\documentclass[ showpacs,twocolumn]{revtex4}
\usepackage[dvips]{epsfig,graphicx}
\usepackage{float,amsfonts}
\usepackage[dvipdfm,colorlinks=true, citecolor=blue, urlcolor=blue ]{hyperref}

\def\be{\begin{equation}}
\def\ee{\end{equation}}
\def\la{\langle}
\def\ra{\rangle}

\begin{document}
\title{Critical Casimir Interactions
Between Spherical Particles in the Presence of the Bulk Ordering Fields.}

\author{O. A. Vasilyev}
\affiliation{Max-Planck-Institut f{\"u}r Intelligente Systeme,
  Heisenbergstra{\ss}e~3, D-70569 Stuttgart, Germany}
\affiliation{ IV. Institut f{\"u}r Theoretische Physik,
  Universit{\"a}t Stuttgart,  Pfaffenwaldring 57, D-70569 Stuttgart, Germany}

\begin{abstract}
The spatial suppression of order parameter fluctuations in a critical media 
produces Critical Casimir forces acting on confining surfaces.
This scenario is realized in a critical binary mixture near the 
demixing transition point that corresponds to the second order phase 
transition of the Ising universality class. 
Due to these critical interactions
similar colloids, immersed in a critical binary mixture near the
consolute point, exhibit attraction.
The numerical method for computation of the 
interaction potential between  two spherical particles
using Monte Carlo simulations for the  Ising model is proposed.
This method is based on the integration of the local magnetization
over the applied local magnetic field. 
For the stronger interaction 
the concentration of the component of the mixture 
that does not wet colloidal particles, should be larger, than the critical 
concentration.  The strongest amplitude of the 
interactions is observed  below the critical point.

\end{abstract}


\maketitle
\section{Introduction}

In 1948 Hendrick Casimir predicted 
that in the vacuum between two parallel perfectly conducting 
plates  an attractive force  appears~\cite{Casimir}. 
This force is caused by the suppression 
 of the zero level quantum fluctuations of the electromagnetic 
 field in the space  between  plates. 
 This phenomena is known as the 
 quantum Casimir effect.

In the  vicinity of the second-order phase
transition in the critical media 
long-range fluctuations of the order parameter  arise.
This phenomenon is observed, {\it e.g.},
 in the critical liquid binary mixture at the demixing point. 
Fisher and de~Gennes predicted~\cite{FdG},
that the confinement of these fluctuations produces  effective forces
acting on confining surfaces. The appearance of forces due
to spatial suppression of fluctuations of an order parameter
in a critical media 
is now known as the 
{\it Critical Casimir} (CC) effect~\cite{Krech,BDT,Gambassi}.

The phenomenon of colloidal particle aggregation in the critical binary mixture
was first reported in~\cite{BE}.
In the planar geometry the CC effect 
for critical binary mixtures  measured experimentally 
 via the influence on the thickness
of the  liquid wetting films~\cite{Fukuto}. In this case the confining
parallel surfaces are substrate-liquid and liquid-vapor
interfaces. Later on,  interaction forces between 
a colloidal particle and a flat substrate were measured 
directly~\cite{nature,PRE1, Nellen}. 
Critical depletion in colloidal suspensions was studied 
experimentally~\cite{BCPP1,BCPP2}.
The colloidal aggregation in microgravity conditions, 
 caused by  CC interaction, was described in Ref~\cite{MKG}.
The  controlled phase transition in colloidal suspension in the 
 critical binary mixture was studied in~\cite{NFHW}. 
 In this article the interaction potential
 between  colloidal particles was extracted from the pair correlation function.
From the experimental point of view CC interactions 
 provide the possibility 
of tuning an interaction between colloidal particles.
By varying the temperature of the binary mixture 
in the vicinity of the consolute point it is possible
to switch on interactions between colloids in controllable and reversible way.

\begin{figure}[t]
\mbox{\includegraphics[width=0.25\textwidth]{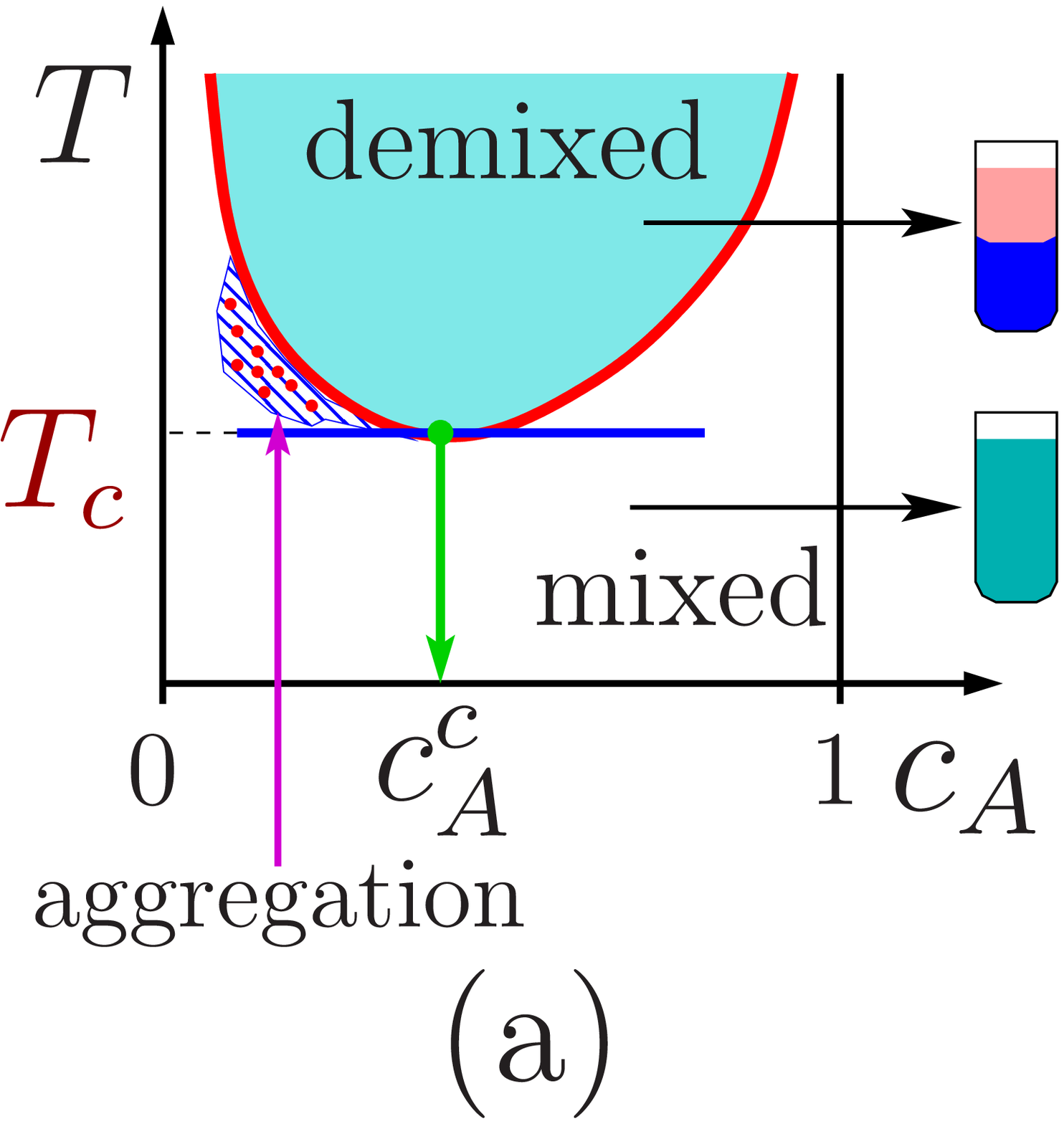}
\includegraphics[width=0.25\textwidth]{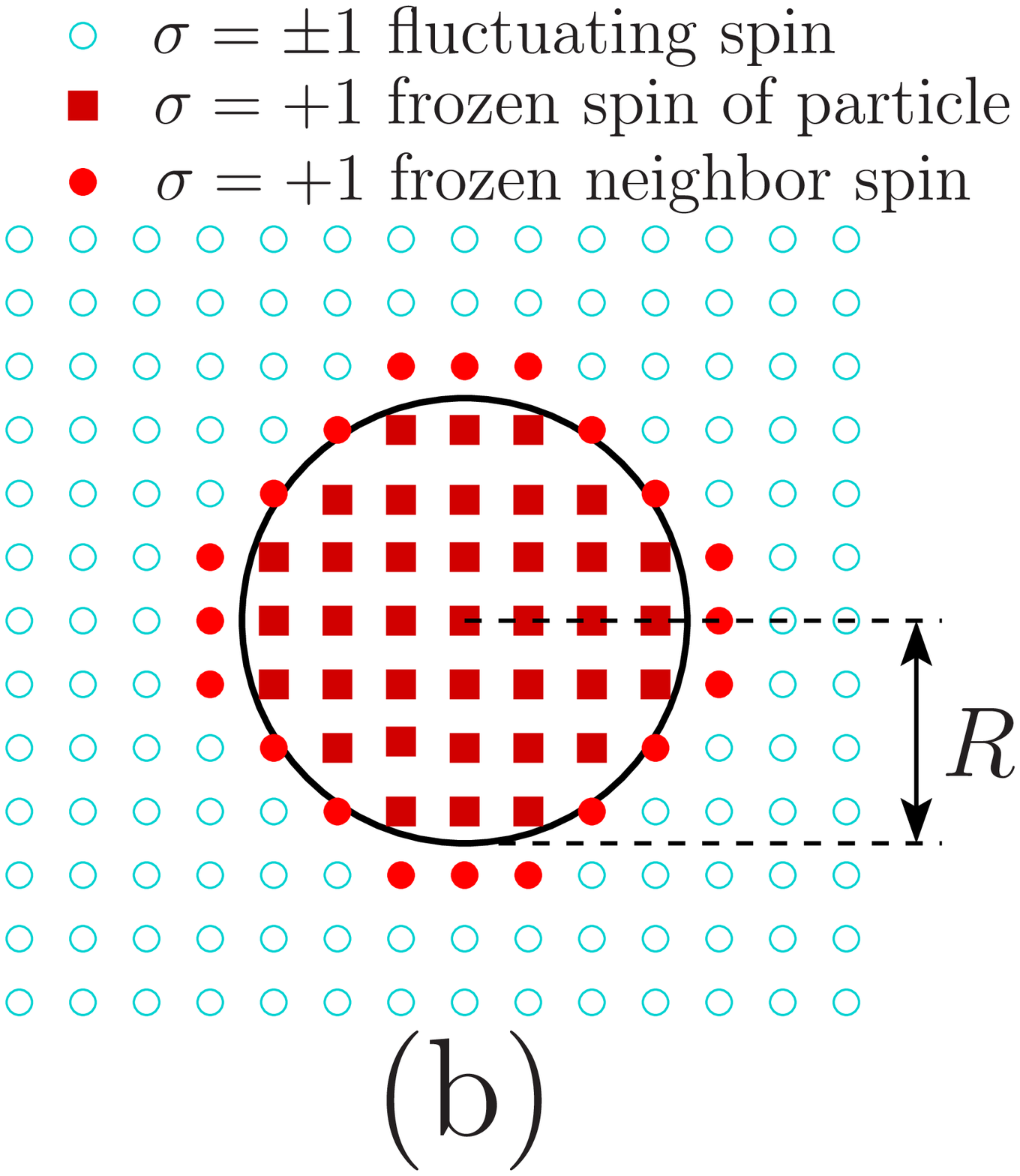}}
\mbox{\includegraphics[width=0.25\textwidth]{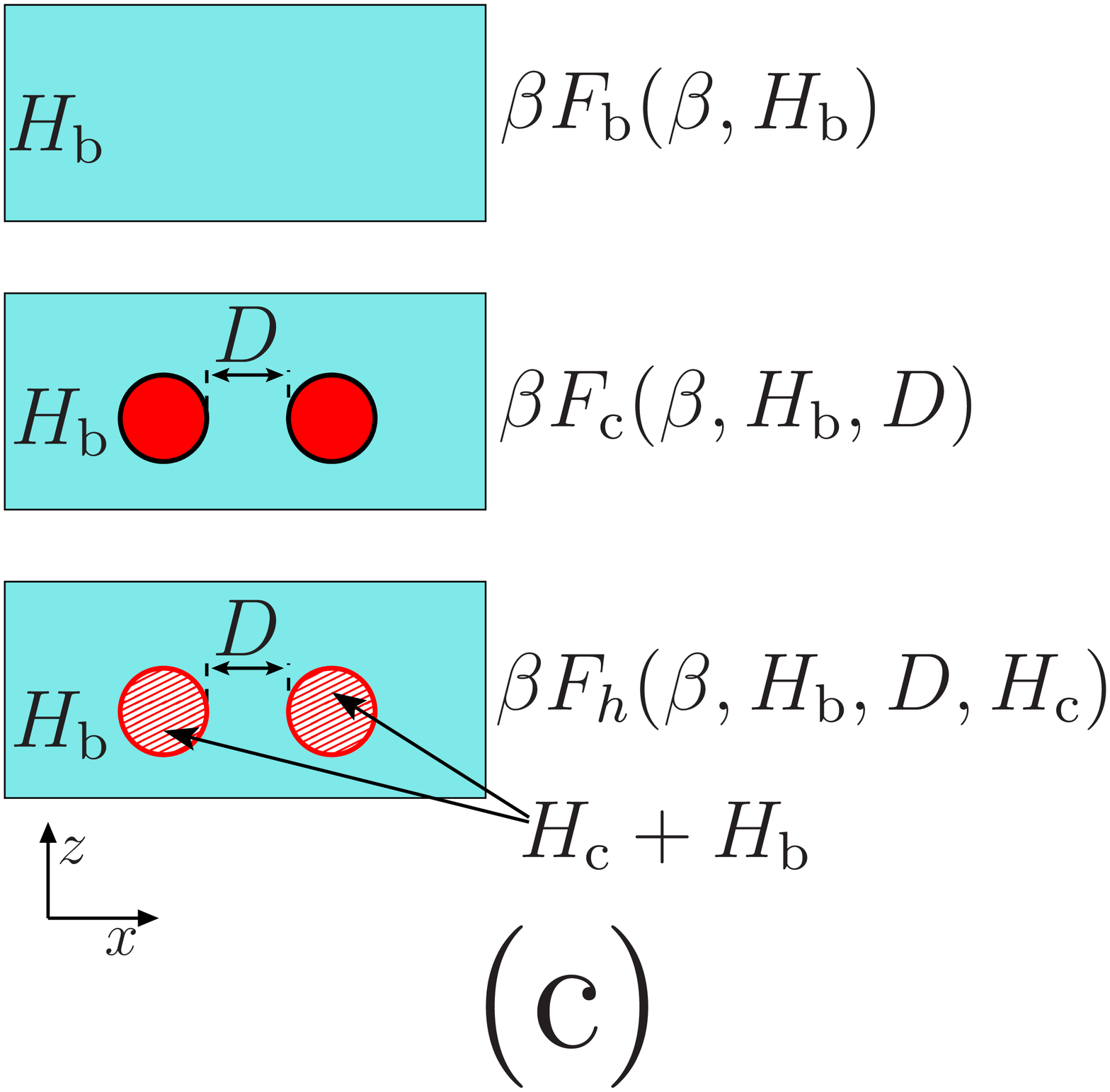}
\includegraphics[width=0.25\textwidth]{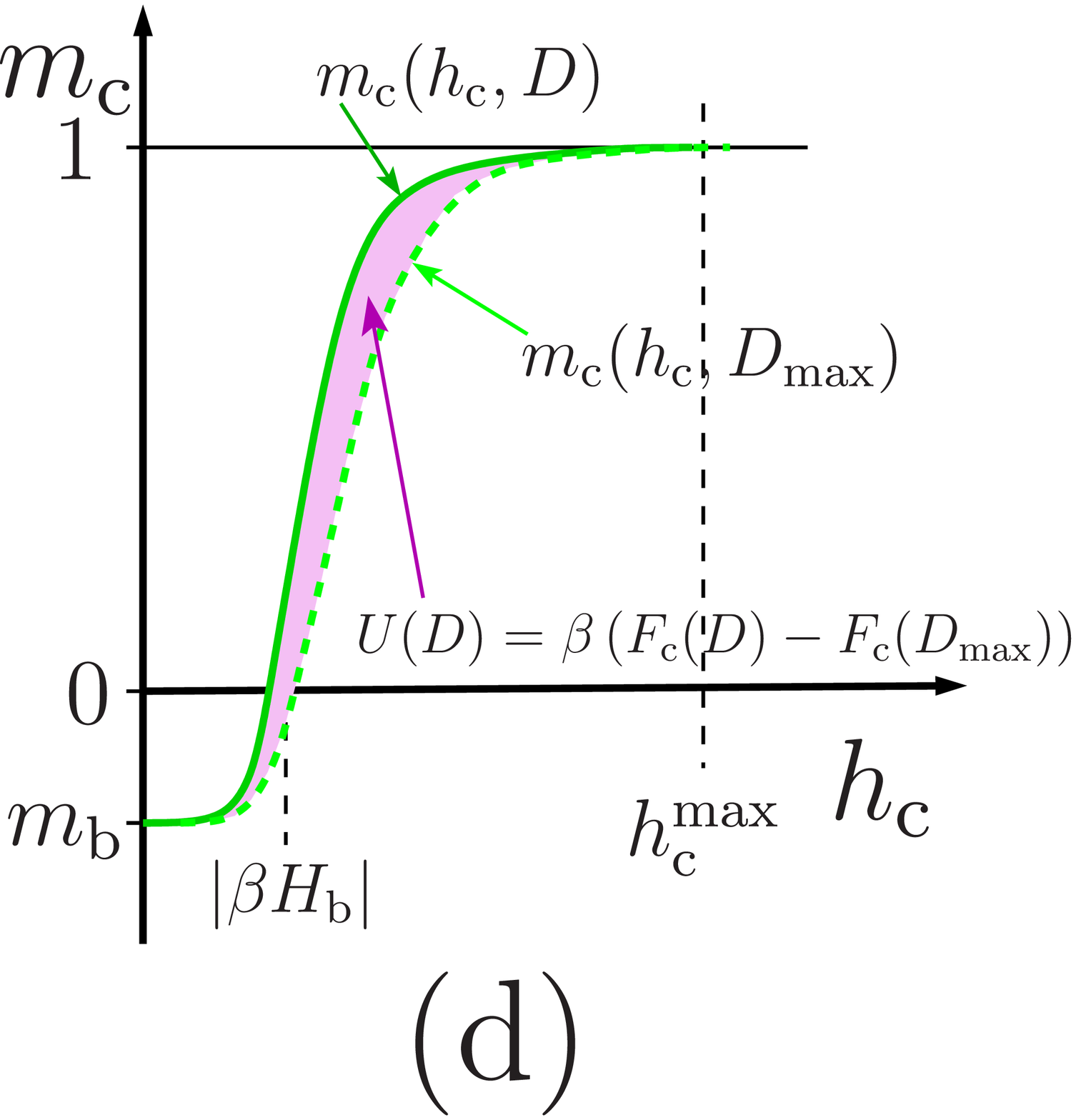}}
\caption{%
(Color online)
(a) Phase diagram of a critical binary mixture 
with the lower critical point and the aggregation region;
(b) Schematic representation for a 
quasi-sphere on the lattice;
(c) Computation of the insertion free energy:
bulk system with the free energy $F_{\mathrm b}(\beta)$, 
the system with fixed spins in two
colloidal particles at distance $D$ with  $F_{\mathrm c}(\beta,H_{\mathrm b},D)$,
the system with  an external 
field $H_{\mathrm c}$ applied to spins of  two
colloidal particles at distance $D$ with the free energy 
$F_{\mathrm h}(\beta,H_{\mathrm b},D, H_{\mathrm c})$, 
(d) Typical graphs of magnetizations 
$m_{\mathrm c}(H_{\mathrm c}, D)$,
$m_{\mathrm c}(H_{\mathrm c}, D_{\mathrm{max}})$
as functions of `colloid' field $h_{\mathrm c}$ 
for separations $D$, $D_{\mathrm{max}}$.
The  shadowed area between curves is equal to the 
absolute value of the 
free energy difference $ U(D)=
\beta F_{\mathrm c}(D)-\beta F_{\mathrm c}(D_{\mathrm{max}})$.
 }
\label{fig:fig}
\end{figure}

The critical binary mixture consists of components 
A and B (with concentrations $c_{A}$ and $c_{B}=1-c_{A}$,
respectively) with the critical concentration $c_{A}^{c}$ and
the critical  temperature $T_{c}$. 
The schematic phase diagram 
with the lower critical point (that corresponds to
the water-lutidine mixture used in experiments~\cite{BE,nature,PRE1, Nellen})
 is shown in Fig.~\ref{fig:fig}(a). 
The state of such a system  is characterized by the 
reduced temperature $t_{AB}=(T-T_{c})/T_{c}$ and 
chemical potentials $\mu_{A}$, $\mu_{B}$ for the two components A and B
with corresponding  values $\mu_{A}^{c}$, $\mu_{B}^{c}$ at criticality.
It is convenient to represent chemical potentials as a combination
of  $H_{AB}=\mu_{A}-\mu_{A}^{c}-(\mu_{B}-\mu_{B}^{c})$ 
(which plays a role of the bulk ordering field)
and $\delta \mu=\mu_{A}+\mu_{B}-(\mu_{A}^{c}+\mu_{B}^{c})$
(which describes the deviation of chemical potential for both components from the critical values).
In the most general case, in the vicinity of the critical point
the state of the binary liquid mixture is characterized by 
two scaling fields that are linear combinations of these three variables
$t_{AB}$, $H_{AB}$, and $\delta \mu$  (see~\cite{Wilding} for detailed description).

A critical binary mixture belongs to the universality class
of the Ising model which state is characterized by the reduced 
temperature $t=(T-T_{c})/T_{c}$ and the bulk magnetic field $H_{\mathrm b}$.
  We consider the  potential difference that is proportional to the  
 bulk field  $H_{AB} \propto H_{\mathrm b}$
   and equal values of the reduced temperatures $t_{AB}=t$. 
 
In accordance with the scaling theory~\cite{Barber,Privman}
the CC interactions are characterized by
the ratio of the linear size of the system 
and the bulk correlation length 
$\xi(t,H_{\mathrm b})$ which is the function of the 
reduced temperature $t$
 and the bulk field $H_{\mathrm b}$. 
  For correct interpretation of experimental results
 we need information about CC interactions of
  colloids for the three-dimensional (3D) Ising universality 
 class. 
 
     The CC force and its scaling function of the 3D Ising universality class
 for the film geometry and various boundary conditions 
 were studied numerically  without the bulk field~\cite{EPL,hucht,PRE,hasen1}.
 Recently, Monte Carlo (MC) simulation results for the plane geometry with the 
 bulk field were obtained~\cite{Myhb,Arxiv_exp_prot}.  
    Results for the CC forces between a spherical particle and a plane 
 for the 3D Ising universality class without 
the bulk field are published in~\cite{Hsp}.
  The CC force between two colloidal particles for the Mean Field (MF) 
universality class was first studied in~\cite{SHD} 
using the conformal transformation.
Without the  bulk field MF interactions between an elliptic  
particle and a wall were studied in~\cite{KHD} and
multi-particle interactions were studied in \cite{THD}. 
  Recently, the results for CC force between two colloidal particles
in the presence of the bulk ordering field  
for the MF universality class were published~\cite{KM}.
       Results for  CC force between two disks 
 for the two-dimensional (2D) Ising model  
 with the bulk field were obtained via
the Derjuaguin approximation~\cite{ZMD}. 
 The alternative method that uses the distribution function 
 of a mobile object position for the computation of the CC interaction 
  was recently proposed~\cite{HH}. In that article the 
  CC interaction potential between a disk and a wall 
  was computed for 2D geometry.

 In the present paper we propose the numerical method for the 
 direct computation of CC interactions between particles
 for a 3D Ising model with the bulk ordering field.
 We  present results for the interaction 
potential  for two particles as a function of the 
bulk field at fixed temperatures  and as functions of the temperature for
fixed values of $H_{\mathrm b}$.
  The paper is organized as follows:  in the second section we describe the numerical 
 method. In the third section results of the MC simulation for the
 interaction energy between two particles are presented. 
 The last section is the conclusion.

\section{Method}
We consider the Ising model on a simple cubic lattice
with periodic boundary conditions,
all distances are measured in lattice units.
The system size is $L_{x} \times L_{y} \times L_{z}$.
In a site $i$ of the lattice the classical
spin $\sigma_{i}= \pm 1$ is located. The inverse temperature
is $\beta=1/(k_{\mathrm B}T)$.
Our aim is to study the interaction between 
 colloidal particles immersed in the critical binary mixture.
Therefore we need the lattice representation of colloidal particles.
The idea proposed by Martin Hasenbusch~\cite{Hsp} 
is to draw a sphere of a certain radius $R$ around a  selected spin. 
Then all spins within the sphere are considered to belong 
to the colloidal particle and fixed to be $+1$. 
In Fig.~\ref{fig:fig}(b)
we plot a cross section of a sphere of the  radius $R=3.5$, 
spins inside the sphere
are denoted by filled squares. We consider the case of 
very strong positive surface fields for colloids. 
This choice corresponds to the symmetry-breaking 
Boundary Conditions (BCs) with completely ordered surface 
and usually denoted as $(++)$ BC~(see~\cite{Surf} for details). 
It means, that a neighbor spin $j$, 
that is in a contact with a particle surface 
will be frozen $\sigma_{j}=+1$; such spins are denoted by filled 
circles. Let us denote as $\{ {\rm col} \}$ the set of all
frozen spins in the system (spins in both colloidal particles
and their neighbors, totally $N_{c}$ spins) and refer to this set as spins of colloidal particles.
 These spins are shown by filled symbols
in  Fig.~\ref{fig:fig}(b). Fluctuating spins in the bulk are denoted  
by empty circles.

Let us denote as  $F_{\mathrm b}$ the free energy 
of an empty bulk system (see Fig.~\ref{fig:fig}(c) top) with 
the standard Hamiltonian 
for a spin configuration $\{ \sigma \}$
\be
\label{eq:ham_b}
{\cal H}_{\mathrm b}(\{ \sigma \})=-J \sum \limits_{\la ij \ra }\sigma_{i} \sigma_{j}
- H_{\mathrm b} \sum \limits_{n }  \sigma_{n},
\ee
where $J=1$ is the interaction constant, $H_{\mathrm b}$
is the bulk magnetic field, the sum $\la ij \ra$
is taken over all neighbor spins,
 the sum over $n$ is taken over all spins of the spin configuration
 $\{ \sigma \}$.
  The free energy of the system is expressed
  via the sum over all possible spin configurations $\Omega$
  as $ F_{\mathrm b}(\beta,H_{\mathrm b})
  =- \frac{1}{\beta}\log \left[ \sum \limits_{\{ \sigma \} \in \Omega } 
  {\mathrm e}^{-\beta {\cal H}_{\mathrm b}(\{ \sigma  \})} \right]$.
The system with two colloidal particles of a radius $R$ at a distance $D$ 
(see Fig.~\ref{fig:fig}(c) middle)
is described by the same Hamiltonian Eq.(\ref{eq:ham_b}).
But all spins $\sigma_{k} \in \{ {\rm col} \} $
 of colloidal particles and their neighbors $\{ {\rm col} \}$
should be frozen   $\sigma_{k}=+1$,
$k \in \{ {\rm col} \} $, so the free energy is
\be
\label{eq:F_c}
 F_{\mathrm c}(\beta,H_{\mathrm b},D)=
- \frac{1}{\beta} 
\log \left[
\sum \limits_{\{ \sigma \} \in \Omega }\prod \limits_{ k  \in \{ {\rm col} \} } 
\delta_{\sigma_{k},1}
  {\mathrm e }^{-\beta {\cal H}_{\mathrm b}(\{ \sigma  \}) } \right].
\ee
Here the product of the Dirac delta functions 
$\delta_{\sigma_{k},1}$ fixes  the values of spins in  colloidal particles
$ k  \in \{ {\rm col} \}$ to be $+1$.
In this expression for a free energy we also
count the interaction between frozen spins within particles.
Let us consider the system with the Hamiltonian 
\be
\label{eq:ham_h}
{\cal H}_{\mathrm  h}(\{ \sigma \})=
-J \sum \limits_{\la ij \ra }\sigma_{i} \sigma_{j}
- H_{\mathrm b} \sum \limits_{n }  \sigma_{n}
- H_{\mathrm c} \sum \limits_{ k \in \{ {\rm col}\} }  \sigma_{k},
\ee
where the additional external local magnetic field $H_{\mathrm c}$
is applied to spins $\sigma_{k}$
of colloidal particles $ k \in  \{ {\rm col}\}$ (see Fig.~\ref{fig:fig}(c)
bottom).
The free energy of this system is given by the formula 
\be
 F_{\mathrm h}(\beta,H_{\mathrm b},D,H_{\mathrm c})
  =- \frac{1}{\beta}\log \left[\sum \limits_{\{ \sigma \} \in \Omega }
  {\mathrm e}^{-\beta {\cal H}_{\mathrm h}(\{ \sigma  \}) } \right].
\ee  
   For zero additional field 
   this free energy equals the free energy of the system without particles
   $ F_{\mathrm h}(\beta,H_{\mathrm b},D,H_{\mathrm c}=0)
= F_{\mathrm b}(\beta,H_{\mathrm b})$. We consider systems
with certain bulk field $H_{\mathrm b}$ at fixed inverse temperature
$\beta$. Therefore in this section we omit arguments 
 $(\beta, H_{\mathrm b})$ of functions for the simplicity of notations.
For a very strong additional field $\beta H_{\mathrm c} \gg 1 $  it has a limit 
$\lim \limits_{\beta H_{\mathrm c} \to \infty }
F_{\mathrm h}(H_{\mathrm c},D)
=F_{\mathrm c}(D)-H_{\mathrm c}N_{\mathrm c} $,
where $N_{\mathrm c}   $ is the total number of spins 
in the colloidal particle  $\{ \rm col \}$, because these 
spins became frozen by the local field $H_{\mathrm c}$.
 Let us introduce the variable $h_{\mathrm c}=\beta H_{\mathrm c}$.
Then the magnetization of spins in colloids 
$M_{\mathrm c}=\sum \limits_{k \in \{ \mathrm{ col} \}}\sigma_{k}$ 
is expressed via the derivative of the free energy 
with respect to $h_{\mathrm c}$:
\be
M_{\mathrm c}(h_{\mathrm c},D)=
-\frac{\partial  
\left[\beta F_{\mathrm h}( h_{\mathrm c}/\beta,D)\right]}
{\partial h_{\mathrm c} }
\ee
Introducing the normalized (per total number $N_{c}$ of spins in particles) 
particle magnetization
$m_{\mathrm c}( h_{\mathrm c},D)=M_{\mathrm c}( h_{\mathrm c},D)/N_{\mathrm c}$,
we can express the free energy via an integral over the magnetization
\be
\beta F_{\mathrm h}(H_{\mathrm c},D)=\beta F_{\mathrm b}- N_{\mathrm c}
\int \limits_{0}^{\beta  H_{\mathrm c}} m_{\mathrm c}(h_{\mathrm c},D)
  {\mathrm d} h_{\mathrm c}.
\ee
Selecting some big enough  maximal value 
of the additional field $h^{\mathrm{max}}_{\mathrm c} \gg 1$
we can express the free energy of the system with colloidal particles as 
\be
\label{eq:Fc_int}
\beta F_{\mathrm c}(D)=\beta F_{\mathrm b}+ N_{\mathrm c} 
\int \limits_{0}^{  h_{\mathrm c}^{\mathrm{max}}}
 \left[1- m_{\mathrm c}(h_{\mathrm c},D)
 \right] {\mathrm d} h_{\mathrm c}.
\ee
The particle magnetization at zero additional field 
$h_{\mathrm c}=0$ equals  the bulk magnetization 
$m_{\mathrm c}(h_{\mathrm c}=0,D)=m_{\mathrm b}$ and it is equal to 1 at
strong $h_{\mathrm c}\gg 1$ field 
$\lim \limits_{h_{\mathrm c} \to \infty}
m_{\mathrm c}( h_{\mathrm c},D)=1$. For this reason
the result of the integration in Eq.(\ref{eq:Fc_int})
does not depend on the upper limit of the integration 
for big enough $ h_{\mathrm c}^{\mathrm{max}}$ 
(we use the value $ h_{\mathrm c}^{\mathrm{max}}=5$). 
In Fig.~\ref{fig:fig}(d) we schematically plot 
the magnetization $m_{\mathrm c}(h_{\mathrm c},D)$
for the case of the negative bulk magnetic field $H_{\mathrm b}<0$.
Graphically, the ``insertion'' free energy 
$\beta F_{\mathrm c}(D)-\beta F_{\mathrm b}$ equals  the area
between lines $m_{\mathrm c}(h,D)$ and 1.

Our final aim is to compute the potential 
$U(D)$ of the Casimir force
$f_{\mathrm C}(D) $
between two quasi-spherical particles at the distance 
$D$ expressed in units $k_{\mathrm B}T$.
  Up to a certain constant $C_{1}$
this potential may be expressed via the free energy
$U(D)=\beta F_{\mathrm c}(D)+C_{1}$. We select this constant 
equal to the value (with the sign ``$-$'') of the  free energy at some maximal separation 
$D_{\mathrm{max}}$: $C_{1}=-\beta F_{\mathrm c}(D_{\mathrm{max}})$.
Therefore $ U(D)=N_{\mathrm c} 
\int \limits_{0}^{  h_{\mathrm c}^{\mathrm{max}}} 
[m_{\mathrm c}(h_{\mathrm c},D_{\mathrm{max}})
-m_{\mathrm c}(h_{\mathrm c},D)]
  {\mathrm d} h_{\mathrm c}.
$
    Graphically, in Fig.~\ref{fig:fig}(d), this function equals the area 
between lines $m_{\mathrm c}(h_{\mathrm c},D)$ and  
$m_{\mathrm c}(h_{\mathrm c},D_{\mathrm{max}})$
with the minus sign.
      This method is optimized for the computation of the
potential of the Casimir interaction $U$.
For the computation of the Casimir force 
 $f_{\mathrm C}=-\frac{\partial  U(D)}{\partial D}$
 between two particles it would be preferable to use 
 the modification of the  proposed method in which we interpolate
between two configurations for distances $D$ and $D-1$ 
 by varying the local field $H_{\mathrm c}$. 

\section{Results} 
\begin{figure}[h]
\includegraphics[width=0.4\textwidth]{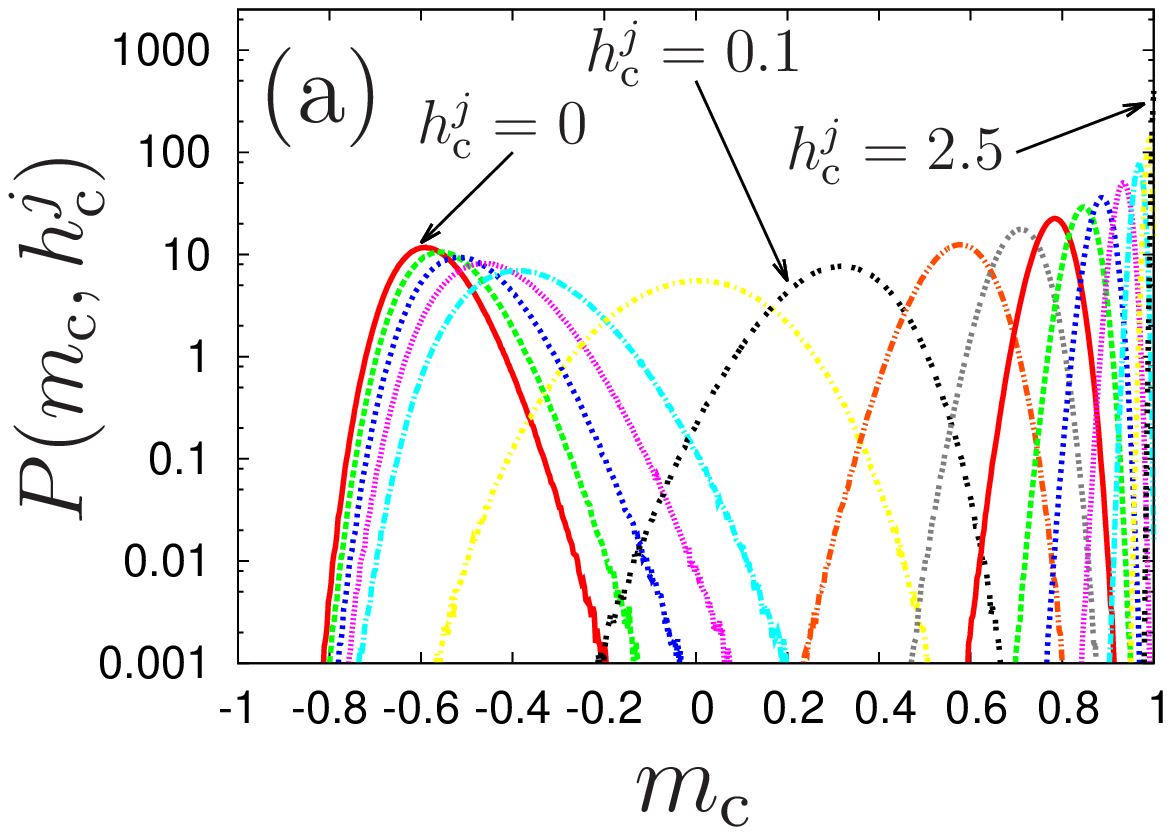}
\includegraphics[width=0.4\textwidth]{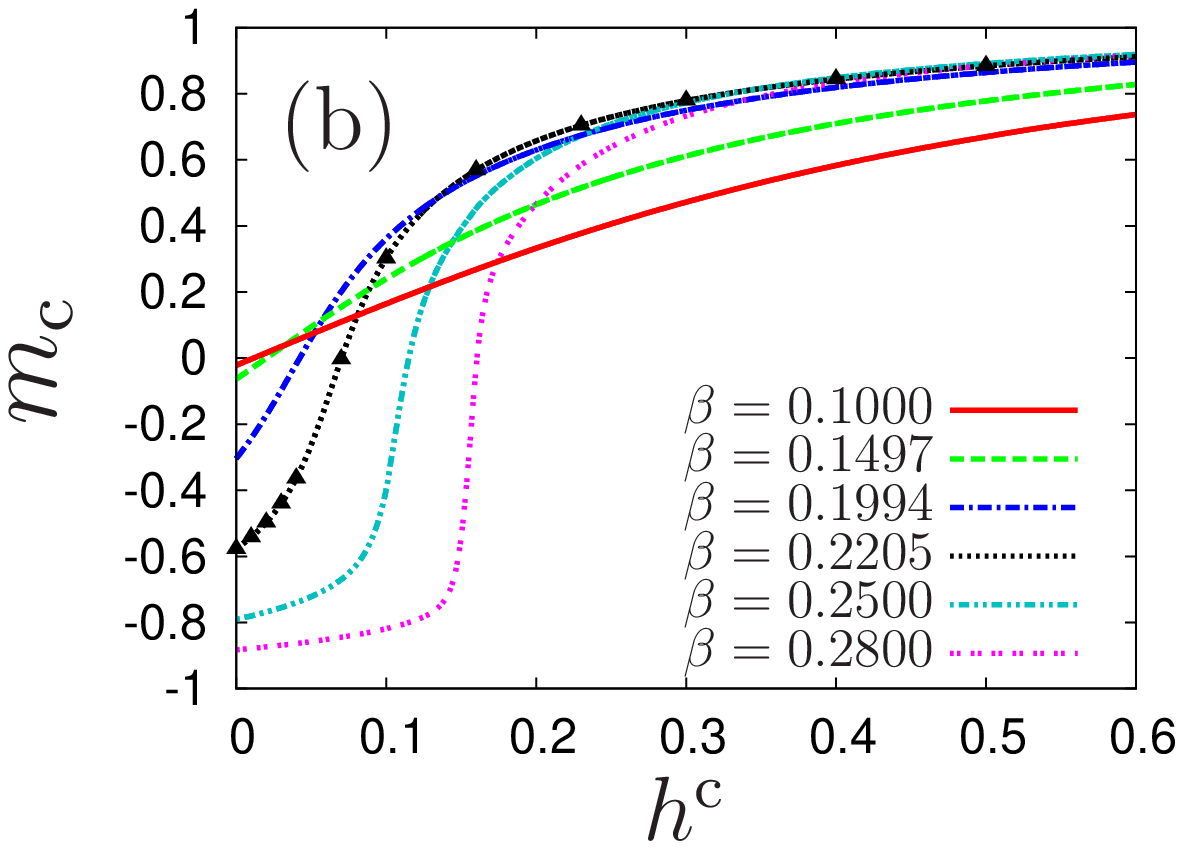}
\caption{%
(Color online) Numerical results for for the 
separation $D=3$ between two spheres of radius
$R=3.5$, the value of the bulk field
$H_{\mathrm b}=-0.1$.
(a) Probability distribution function $P(m_{\mathrm c},h_{\mathrm c}^{j})$
of the magnetization $m_{\mathrm c}$  for the inverse temperature 
$\beta=0.2205$
and various values of the local field (from left to right)
$h_{\mathrm c}^{j}=0,0.01,\dots,2.5$.
(b) Average particle magnetization $m_{\mathrm c}$
as a function of the local field $h_{\mathrm c}$
for various
values of the inverse temperature 
$\beta=0.1,0.1497,0.1994,0.2205,0.25,0.28$;
black triangles correspond to lines from  panel~(a).
 }
\label{fig:pm}
\end{figure}
We perform numerical simulations for the system 
of the size $78 \times 49 \times 49$. Two
quasi-spherical particles of the radius $R=3.5$
are located at separation $D$ along the $x$ direction
(see the $x-z$ cross section in Fig.~\ref{fig:fig}(c)).
For separation $D=0$ the particles are in the contact.
The  separation $D_{\mathrm{max}}=30$ is the maximal
possible interparticle separation in $x$ direction for this system.
For accurate integration over the particle magnetization we use the histogram 
 reweighting technique~\cite{FS,LB}.
 The probability
 distribution $P(m_{\mathrm c}, h_{\mathrm c})$ of the particle magnetization $m_{\mathrm c}$
is proportional to the exponent 
$P(m_{\mathrm c}, h_{\mathrm c}) 
\propto \exp( h_{\mathrm c} N_{\mathrm c} m_{\mathrm c})$.
We compute this probability distribution for 16 values of the additional field
$ h_{\mathrm c}^{j}=\{$0,0.01,0.02,0.03,0.04,
0.05,0.07,0.1,0.16,0.23,0.4,0.5,0.7,1,1.5,2.5$\}$.
The probability distribution for the value of the field  $h_{\mathrm c}$
may be expressed as 
\be
P(m_{\mathrm c}, h_{\mathrm c}) =
\frac{1}{A} \exp[ (h_{\mathrm c}-h_{\mathrm c}^{j} )N_{\mathrm c} m_{\mathrm c}],
\ee
where the normalization constant $A=\sum \limits_{m_{\mathrm c}}
\exp[ (h_{\mathrm c}-h_{\mathrm c}^{j} )N_{\mathrm c} m_{\mathrm c}]$
and the values of fields should be close enough 
to let the probability distributions  intersect.
In Fig.~\ref{fig:pm}(a) we plot the probability
$P(m_{\mathrm c}, h_{\mathrm c}^{j})$ as a function of $m_{\mathrm c}$
for the set of reference points $h_{\mathrm c}^{j}$ for 
$H_{\mathrm b}=-0.1$,  $\beta=0.2205$. In Fig.~\ref{fig:pm}(b)
we plot the magnetization $m_{\mathrm c}$
as a function of $h_{\mathrm c}$ for $H_{\mathrm b}=-0.1$
and various values of $\beta=0.1,0.1497,0.1994,0.2205,0.25,0.28$.
For the curve for $\beta=0.2205$ we denote by triangles values
$ h_{\mathrm c}^{j}$, for which the distribution in  Fig.~\ref{fig:pm}(a)
is computed.

In accordance with the scaling concept  the CC interactions 
 between two similar colloidal particles
of  radius $R$ at   distance $D$ at  temperature $T$,
and the value of the bulk field $H_{\mathrm b}$
are characterized by three variables:
$R$, $D$  and the bulk correlation length
$\xi=\xi(t,H_{\mathrm b})$. Here
 $t=(T-T_{c})/T_{c}=(\beta_{c}-\beta)/\beta$ 
is the reduced temperature ($\beta=1/(k_{B} T)$ is the inverse temperature).
 For the 3D Ising model the value of the critical inverse
  temperature is $\beta_{c}=0.2216544(3)$~\cite{RZW}.
In the general case the correlation length 
is an unknown function of the reduced temperature $t$ and 
of the bulk field $H_{\mathrm b}$.
But for zero magnetic field the correlation length is
$\xi_{t}(t) \equiv \xi(t,0)=\xi^{\pm}_{0} t^{-\nu}$ and  
at the critical temperature the correlation length is 
$\xi_{h}(H_{\mathrm b}) \equiv \xi(0,H_{\mathrm b})=
\xi^{H}_{0} |H_{\mathrm b}|^{-\frac{\nu}{\Delta}}$
 where the value of the correlation length 
critical exponent is $\nu=0.63002(10)$~\cite{Hasnu}, $\Delta=1.5637(14)$~\cite{PV} 
and critical amplitudes are $\xi^{H}_{0}=0.3048(3)$~\cite{EFS},
$\xi^{-}_{0}=0.243(1)$, and $\xi^{+}_{0}=0.501(2)$~\cite{RZW}.
\begin{figure*}[t]
\includegraphics[width=0.4\textwidth]{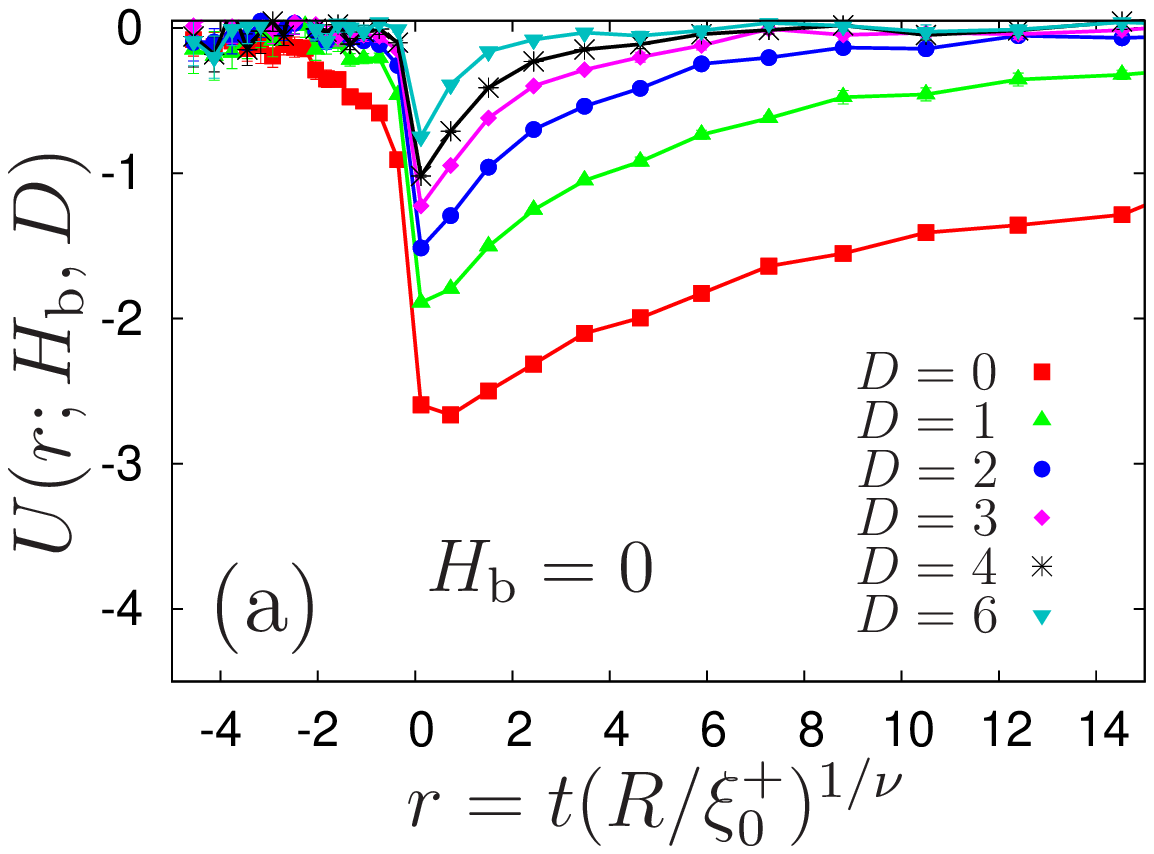}
\includegraphics[width=0.4\textwidth]{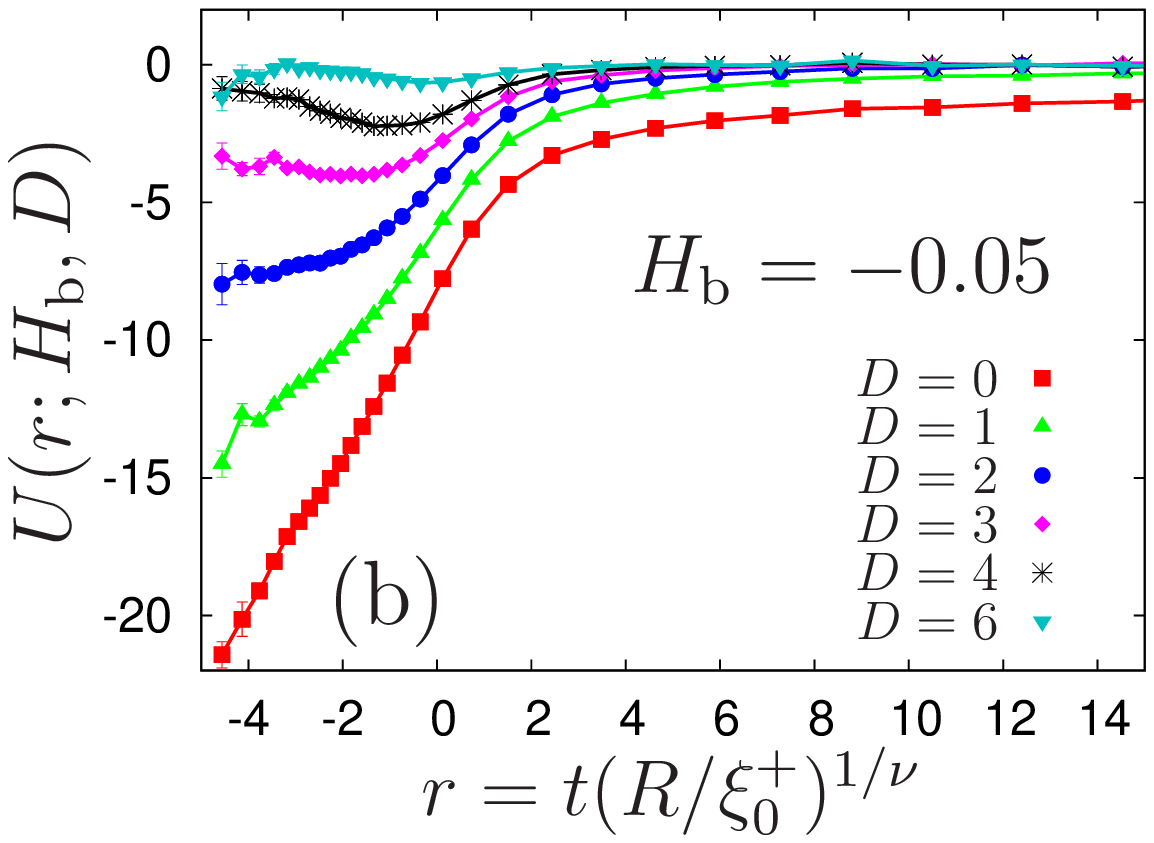}
\includegraphics[width=0.4\textwidth]{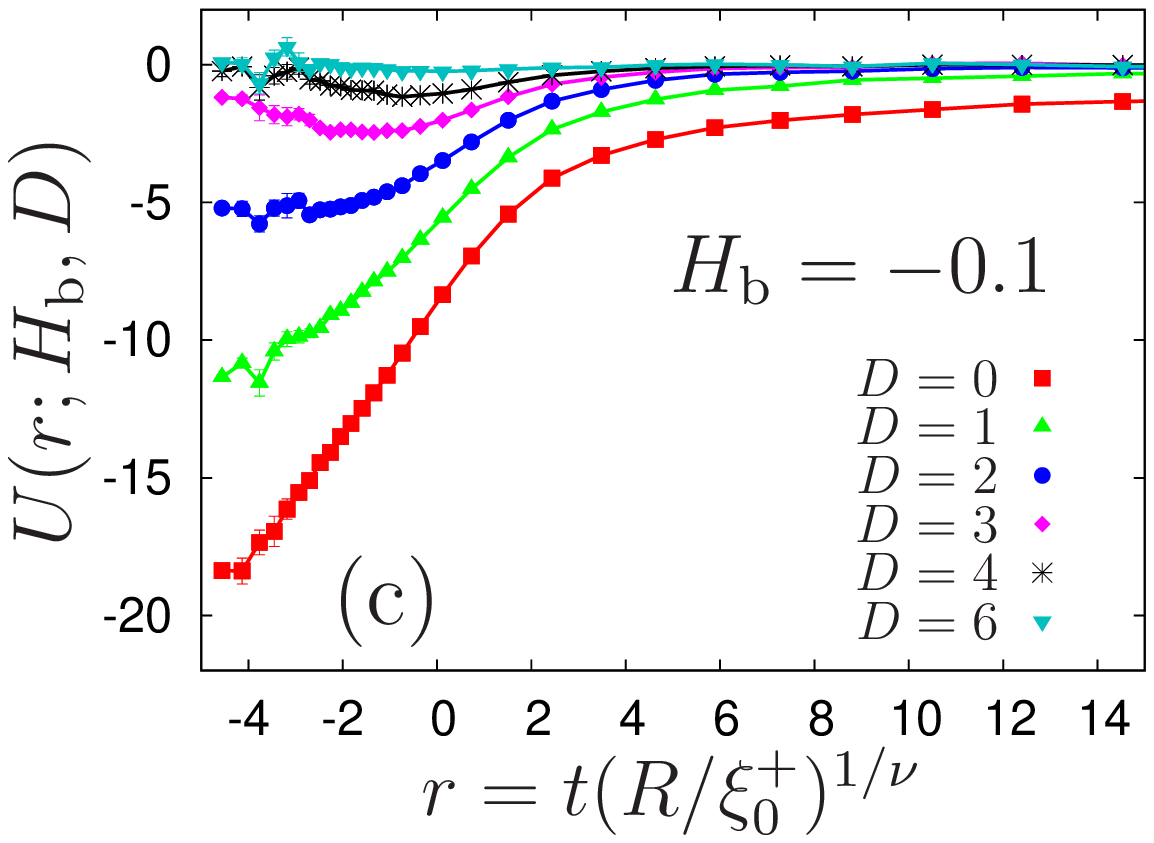}
\includegraphics[width=0.4\textwidth]{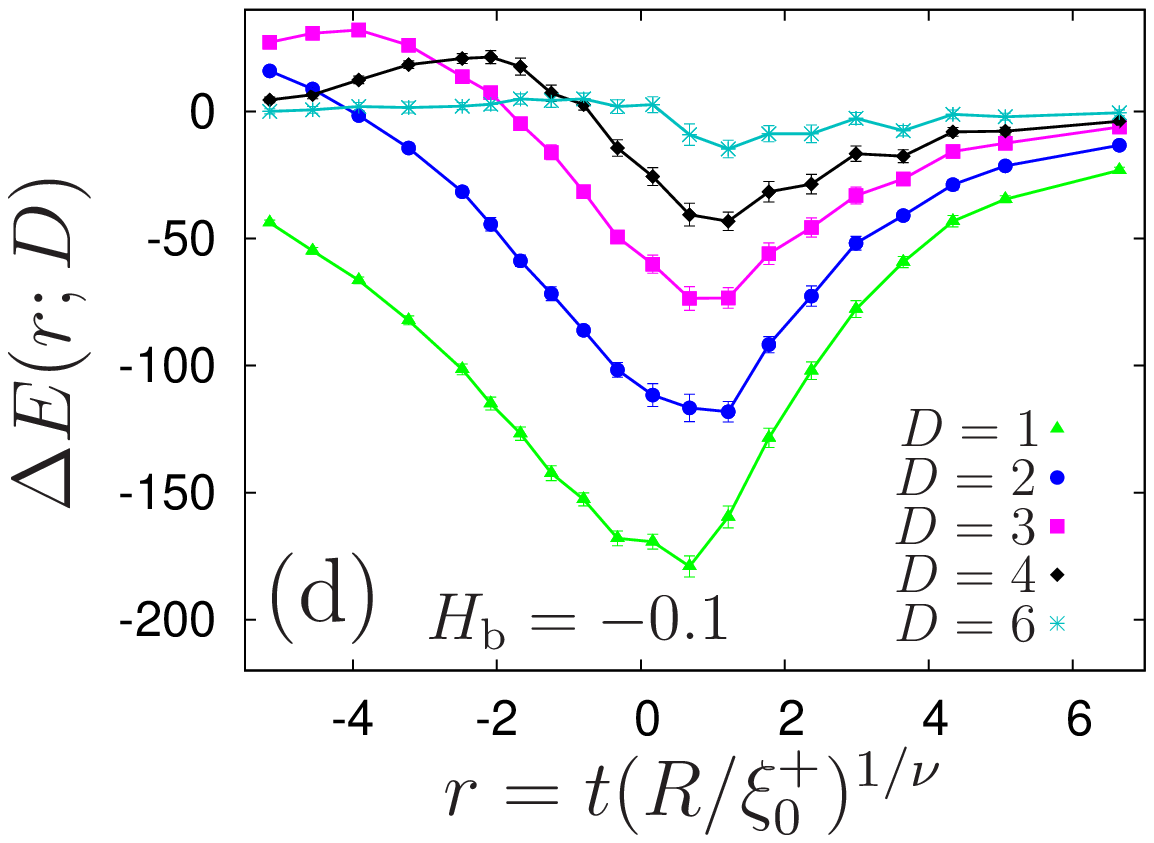}
\caption{%
(Color online)
The Casimir interaction potential $ U(r;H_{\mathrm b},D)$
as a function of the variable  $r= t (R/\xi_{0}^{+})^{1/\nu} $
for  various values of the separation $D=0,1,2,3,4,6$
for: (a) zero bulk field $H_{\mathrm b}=0$,
(b) negative bulk field $H_{\mathrm b}=-0.05$,  
and (c) negative bulk field $H_{\mathrm b}=-0.1$.
(d) The energy difference $\Delta E=E(D)-E(D_{\mathrm{max}})$ as 
a function of the scaling variable   $r $ for $H_{\mathrm b}=-0.1$.
 }
\label{fig:U_H}
\end{figure*}
In the present paper we study two cases: the constant magnetic field
and various temperatures and constant  temperatures and various    
values of the magnetic field.
In the first case we choose the scaling variable 
$r= t (R/\xi_{0}^{+})^{1/\nu}$ 
as an argument of the function because in the case 
of the variable $t (D/\xi_{0}^{+})^{1/\nu}$ for different values of $D$
 we should perform computations for different temperatures
(an alternative choice is the ratio $ \mathrm{sgn}(t)R/\xi_{t}$,
for this scaling variable
 the function is ``stretched'' in the vicinity of zero).
 The second reason for this choice is that
it let us include the distance $D=0$ (when particles touch each other)
into consideration. In the presence of the bulk ordering field,
 critical fluctuations on the system size scale should be suppressed,
therefore in the present paper  we do not study
the influence of the system size.

In Figs.~\ref{fig:U_H}(a)-\ref{fig:U_H}(c) we plot the interaction potential
$ U(r;H_{\mathrm b},D)$ as a function of the scaling variable 
$r$ for separations $D=0,1,2,3,4,6$
and values of the bulk field $H_{\mathrm b}=0,-0.05,-0.1$, respectively.
In the case of zero bulk field Fig.~\ref{fig:U_H}(a),
the attractive potential has a pronounced minimum in the vicinity 
of the critical point $r \simeq 0$. For the negative value of the bulk
field $H_{\mathrm b}=-0.05$ the amplitude of the attractive interaction
increases several times. For big enough separations $D=4,6 >R$
the width of the interaction potential well with respect to $r$
becomes very big. For shorter separations $D=1,2,3 <R$ the minimum 
of the interaction disappears and the interaction within 
the investigated range $-4 < r < 14$ has no minimum.  
The strongest interaction corresponds to the smallest value of $r$.
In Fig.~\ref{fig:U_H}(d) we plot the energy difference 
$\Delta E(r;D,H_{\mathrm b})=E(r;D,H_{\mathrm b})-E(r;D_{\mathrm{max}},H_{\mathrm b})$
as a function of $r$ for separations $D=1,2,3,4,6$ with respect to maximal separation
$D_{\mathrm{max}}=30$
(the same maximal separation is used for the computation of the interaction potential 
 $ U(r;D,H_{\mathrm b})=
\beta F_{\mathrm c}(r;D,H_{\mathrm b})- 
F_{\mathrm c}(r;D_{\mathrm{max}},H_{\mathrm b})$~).
\begin{figure*}[h]
\includegraphics[width=0.4\textwidth]{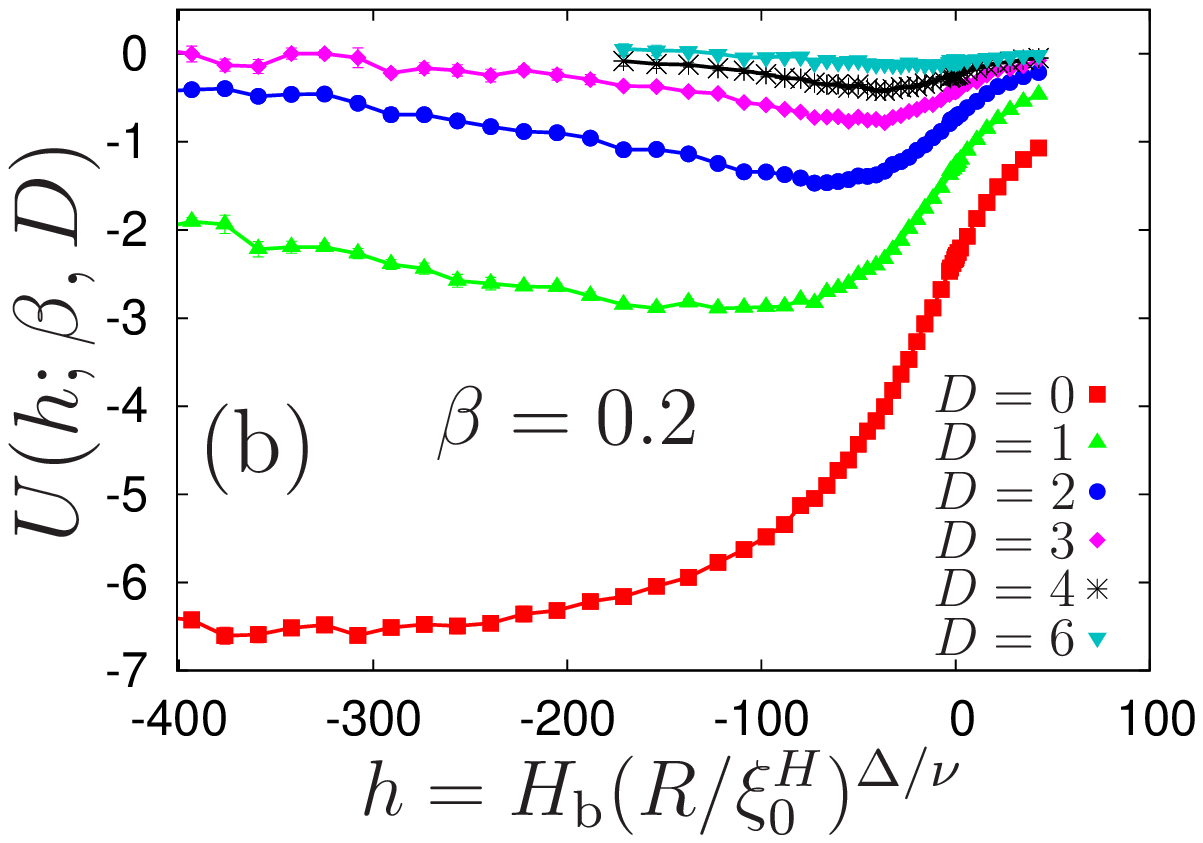}
\includegraphics[width=0.4\textwidth]{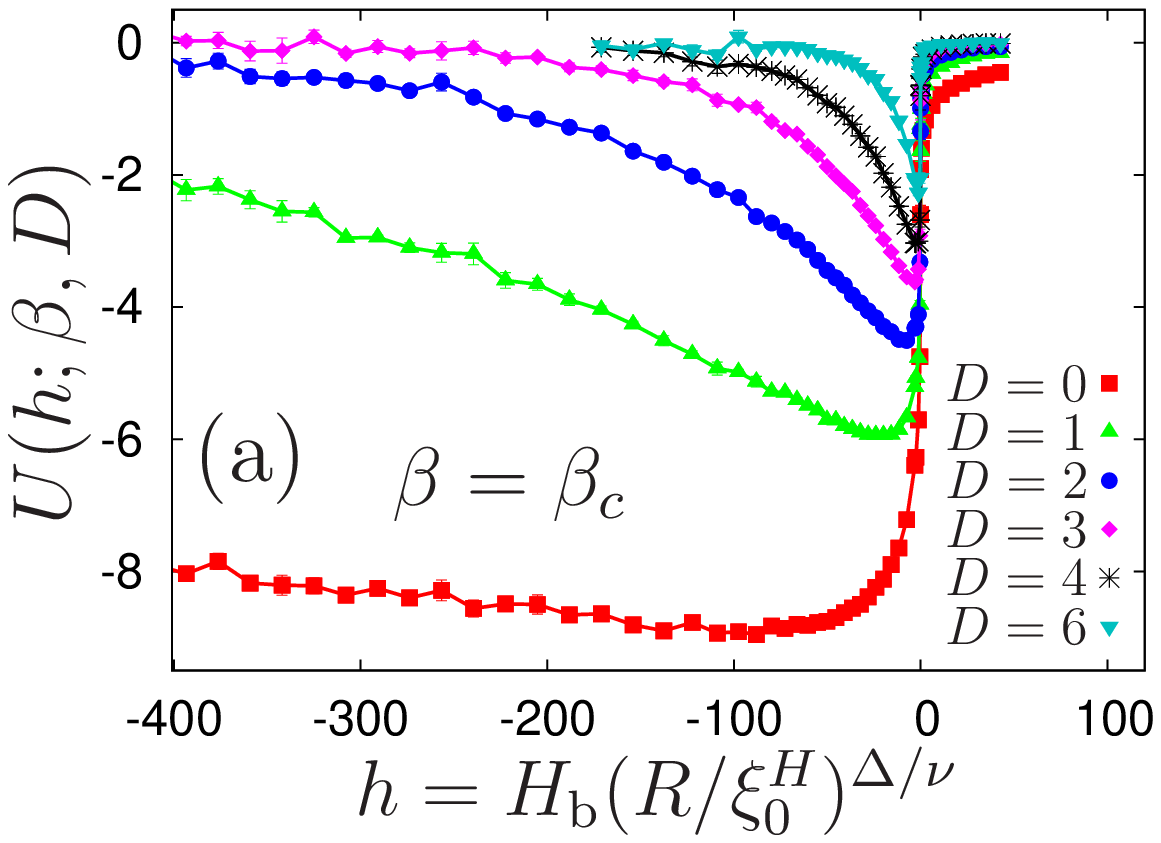}
\includegraphics[width=0.4\textwidth]{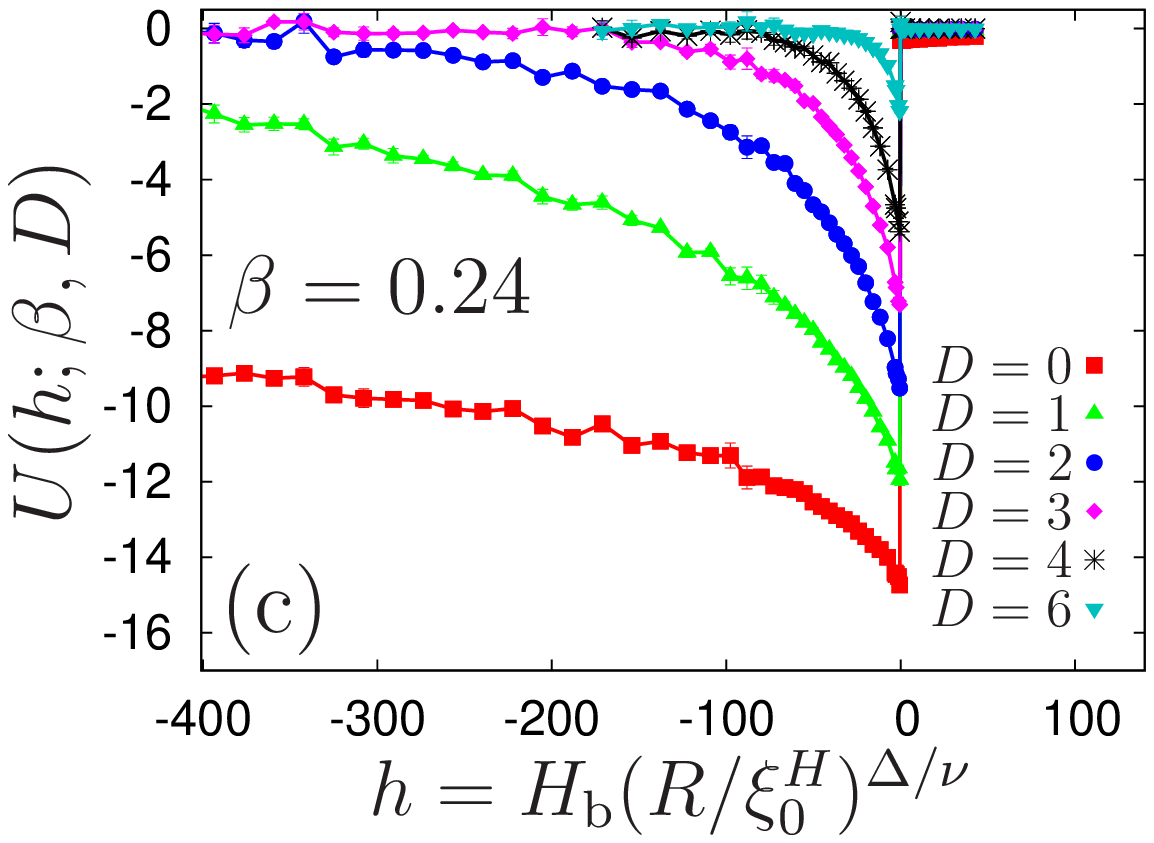}
\includegraphics[width=0.4\textwidth]{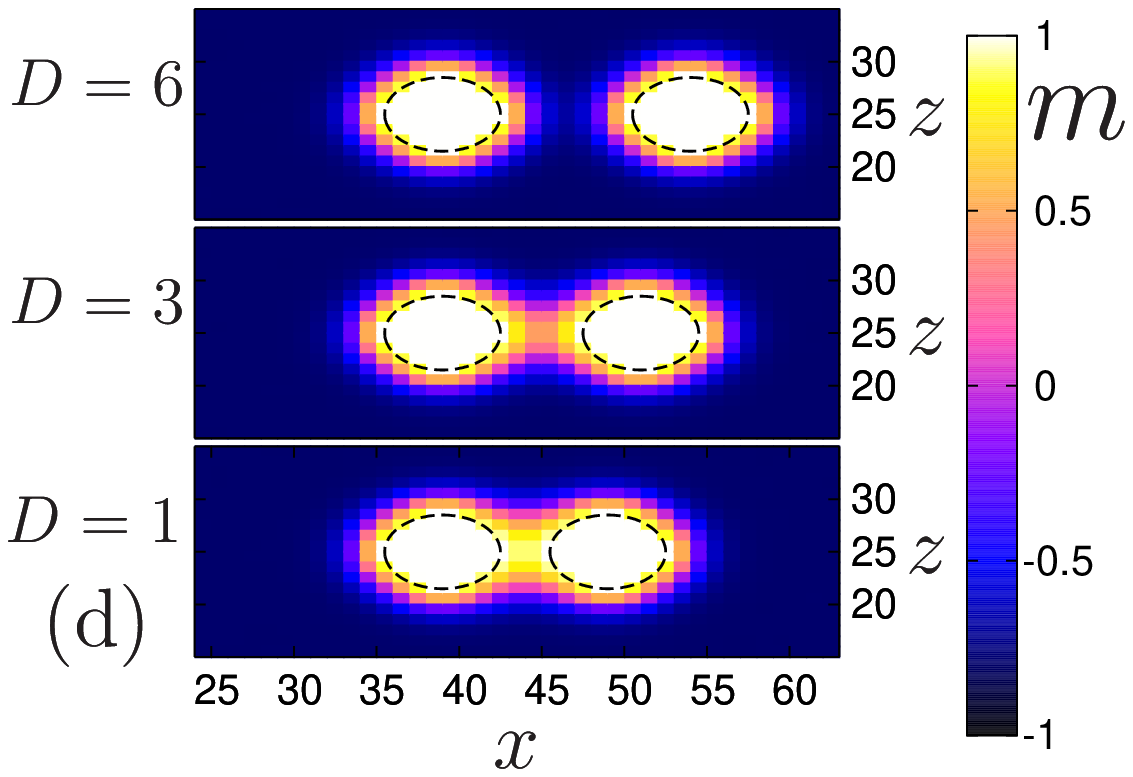}
\caption{%
(Color online)
The Casimir interaction potential $ U(h;\beta,D)$
as a function of the bulk field scaling variable 
$h=H_{\mathrm b} (R/\xi_{0}^{H})^{\Delta/\nu}$
for  various values of separation $D=0,1,2,3,4,6$: 
(a) above the critical  point $\beta=0.2$,
(b) at the critical point $\beta_{c}\simeq 0.221654$,
and (c) below the critical point  $\beta=0.24$.  
(d) The magnetization profile $m(x,z)$ as a function
of coordinates $x,z$ for $\beta=0.25$ ($r\simeq -2.48$),  $H_{\mathrm b}=-0.1$ 
($h\simeq -42.6$), and various separations $D=1,3,6$.
 }
\label{fig:U_beta}
\end{figure*}
In Figs.~\ref{fig:U_beta}(a)-\ref{fig:U_beta}(c) we plot 
the CC interaction potential $U(h;\beta,D)$
as a function of the scaling variable $h=H_{\mathrm b} (R/\xi_{0}^{H})^{\Delta/\nu}$
for various separations and temperatures $\beta=0.2,\beta_{c},0.24$ (above $T_{c}$,
at $T_{c}$, and below $T_{c}$, respectively). 
In Fig.~\ref{fig:U_beta}(d)
we plot the magnetization profile $m(x,z)$ as a function of coordinates $(x,z)$
for the value of the inverse temperature $\beta=0.25$ (the corresponding value of the
scaling variable $r\simeq -2.48$)
and the value  of the magnetic field $H_{\mathrm b}=-0.1$
 (the value of the scaling 
variable $h\simeq -42.6$) using the colormap.
We observe, that for $D>0$ the interaction potential 
has a minimum as a function of $h$. The depth of this minimum decreases with increasing
separation $D$. Above $T_{c}$ the minimum is smooth and is shifted 
for stronger negative values of $h \sim -50,-100$.
Below $T_{c}$ the minima become sharp and narrow, shifted to smaller
(in the amplitude) values of the negative field $  -20<h<0$.
In Fig.~\ref{fig:U_beta}(d)  
we observe for $H_{\mathrm b}=0.1$, $\beta=0.25$ 
($h\simeq -42.6$, $r\simeq -2.48$)
the formation of the bridge
of positive spins  (which corresponds to  component 
$A$ of the binary mixture)
for small separations $D=1,3$. 
For larger separation $D=6$ the bridge disappears.
That correlates with the presence of an attractive potential 
in Fig.~\ref{fig:U_beta}(c) for $h \simeq -40$ and $D=1,3$
and the absence of attraction for $D=6$.  It means, that 
the strong attraction for $D=1,2,3$ in 
 Figs.~\ref{fig:U_H}(b) and~\ref{fig:U_H}(c) for $r <-4$
and in Fig.~\ref{fig:U_beta}(c) for $h <-100$ 
is produced by the formation
of the bridge of positive spins. 
This is confirmed by the energy difference $\Delta E$
in Fig.~\ref{fig:U_H}(d), which has the noticeable minimum for $D=1,2,3$.
It corresponds to the total decreasing of the area of the $-+$ interface
below $T_{c}$ due to the formation of the bridge.  For $D=6$
the energy difference has no  pronounced minimum, in this case the bridge is absent.

\section{Conclusion}

  A numerical method for the computation of the  potential of the CC
interaction between particles immersed in the critical media is proposed.
This method provides results for the  3D Ising universality class
in the presence of non-zero bulk ordering field.
 The potential energy difference for two interparticle 
distances $D$ and $D_{\mathrm{min}}$
 has a simple graphical representation and is
proportional to the area between graphs of the local magnetization 
for these two separations. We compute the interaction 
potential as a function of the temperature scaling variable for  fixed values 
of the bulk ordering field and vice versa, as a function of the
bulk field scaling variable for fixed temperatures.
The strongest interaction for particles 
with $(+)$ boundary conditions 
(for colloidal particles with the surface 
that has a preference to  component A) is observed for 
negative bulk fields $H_{\mathrm b}<0$ (B-rich phase of the binary mixture)
below the critical point $T<T_{c}$ (above the lower critical point 
in the phase diagram Fig.~\ref{fig:fig}(a)). 
This aggregation region is shown in Fig.~\ref{fig:fig}(a)
(as observed in~\cite{BE}).
 For a small interparticle  distances we observe the 
formation of the bridge of $+$ phase between particles 
that produces forces acting 
far away from criticality. 
 As a result of the computation the interaction potential
 between two  colloidal particles is provided
that is convenient for comparison 
with experimental results~\cite{nature,NFHW}.
The proposed method may be applied also to studying 
multi-particle interactions (which play a significant role in the critical 
aggregation in the vicinity of the critical point~\cite{DVN}) 
in a critical solvent.


\clearpage
\begin{thebibliography}{99}

\bibitem{Casimir} H. B. Casimir,
 Proc. K. Ned. Akad. Wet. {\bf 51} 793 (1948).

\bibitem{FdG} 
M. E. Fisher and P. G. de Gennes, C. R. Acad. Sci. Paris Ser. B {\bf 287}, 207
(1978).



\bibitem{Krech} M. Krech, {\it Casimir Effect in Critical Systems}
  (World Scientific, Singapore, 1994).


\bibitem{BDT}  J. G. Brankov, D. M. Dantchev, and N. S. Tonchev, {\it The
    Theory of Critical Phenomena in Finite-Size Systems - Scaling and Quantum
    Effects} (World Scientific, Singapore, 2000).


\bibitem{Gambassi} A. Gambassi, J. Phys.: Conf. Ser. {\bf 161}, 012037 (2009). 

\bibitem{BE} D. Beysens and D. Est\`eve,
Phys. Rev. Lett. {\bf 54}, 2123 (1985). 

\bibitem{Fukuto} M. Fukuto, Y. F. Yano, and  P. S. Pershan, Phys. Rev. Lett.
{\bf 94}, 135702 (2005).

 
\bibitem{nature}
C. Hertlein, L. Helden, A. Gambassi, S. Dietrich, and C. Bechinger, Nature
{\bf 451}, 172 (2008).

\bibitem{PRE1}
A. Gambassi,  A. Macio\l ek,  C. Hertlein, U. Nellen, L. Helden,  C. Bechinger, 
and S. Dietrich,
Phys. Rev. E {\bf 80}, 061143 (2009).

\bibitem{Nellen} U. Nellen, L. Helden, and C. Bechinger,
EPL {\bf 88}, 26001 (2009).

\bibitem{BCPP1} S. Buzzaccaro, J. Colombo, A. Parola, and R. Piazza,
Phys. Rev. Lett. {\bf 105}, 198301 (2010).

\bibitem{BCPP2}
R. Piazza, S. Buzzaccaro, A. Parola, and J. Colombo,
 J. Phys.: Condens. Matter {\bf 23}, 194114 (2011).

 
\bibitem{MKG}  S.J. Veen, O. Antoniuk, B. Weber, M.A.C. Potenza, S. Mazzoni, 
 P. Schall, and G.H. Wegdam,
 Phys. Rev. Lett. {\bf 109}, 248302 (2012). 

\bibitem{NFHW}  V.D. Nguyen,
S. Faber, Z. Hu,   G.H. Wegdam, and  P. Schall,
Nature Comm. {\bf 4},   1584 (2013).

\bibitem{Wilding} N.B. Wilding, Phys. Rev. E {\bf 55}, 6624 (1997).

\bibitem{Barber} M. N. Barber, in {\it Phase Transitions and Critical Phenomena},
edited by C. Domb and J. L. Lebowitz (Academic, New York, 1983), Vol. 8,
p. 149.

\bibitem{Privman} V. Privman, in {\it Finite Size Scaling and Numerical Simulation of
  Statistical Systems}, edited by V. Privman (World Scientific, Singapore,
1990), p. 1.

\bibitem{EPL} 
O. Vasilyev, A. Gambassi, A. Macio\l ek, and S. Dietrich, EPL {\bf 80}, 60009
(2007).

\bibitem{hucht} A. Hucht, Phys. Rev. Lett. {\bf 99}, 185301 (2007).

\bibitem{PRE}
O. Vasilyev, A. Gambassi, A. Macio\l ek, and S. Dietrich,
Phys. Rev. E {\bf 79}, 041142 (2009).


\bibitem{hasen1} M. Hasenbusch, Phys. Rev. B {\bf 82}, 104425 (2010).

\bibitem{Myhb} O. Vasilyev and  S. Dietrich, EPL {\bf 104},
60002 (2013). 

\bibitem{Arxiv_exp_prot} D.L. Cardozo, H. Jacquin, P.C.W. Holdsworth,
preprint  arXiv:1404.4747 (2014). 


\bibitem{Hsp} M. Hasenbusch, Phys. Rev. E {\bf 87}, 022130 (2013).

\bibitem{SHD}  F. Schlesener, A. Hanke, and S. Dietrich, 
J. Stat. Phys. {\bf 110}, 981 (2003).

\bibitem{KHD} S. Kondrat, L. Harnau, and S. Dietrich, J. Chem. Phys. {\bf 131},
234902 (2009).

\bibitem{THD}  T. G. Mattos, L. Harnau, and S. Dietrich, J. Chem. Phys.
{\bf 138 }, 074704 (2013).

\bibitem{KM}  T.F. Mohry,  S. Kondrat, A. Macio\l ek, and S. Dietrich,
preprint  arXiv:1403.5492 (2014).

\bibitem{ZMD}  M. Zubaszewska,  A. Macio\l ek, and A. Drzewi\'nski,
 Phys. Rev. E {\bf 88}, 052129  (2013).

\bibitem{HH} H. Hobrecht and A. Hucht, EPL {\bf 106}, 56005 (2014).
 
\bibitem{Surf}
O. Vasilyev, A. Macio\l ek, S. Dietrich
Phys. Rev. E {\bf 84}, 041605 (2011).

\bibitem{FS} A. M. Ferrenberg and R. H. Swendsen,
Phys. Rev. Lett. {\bf 63}, 1195  (1989).  

\bibitem{LB} D. P. Landau and K. Binder, {\it A Guide to Monte Carlo
  Simulations in Statistical Physics} (Cambridge University Press, London,
  2005).

\bibitem{RZW} C. Ruge, P. Zhu, and F. Wagner, Physica A {\bf 209}, 431 (1994).

\bibitem{Hasnu} M. Hasenbusch, Phys. Rev. B {\bf 82}, 174433 (2010).

\bibitem{PV} A. Pelissetto  and E. Vicari, Phys. Rep. {\bf 368}, 549 (2002).


\bibitem{EFS} J. Engels, L. Fromme, and M. Seniuch,
Nucl. Phys. B {\bf 655}, 277 (2003).

\bibitem{DVN}  M.T. Dang, A.V. Verde, 
    V.D. Nguyen, P.G. Bolhuis, and P. Schall,
    J. Chem. Phys. {\bf 139}, 094903 (2013). 
    \end{thebibliography}
\end{document}